# Room-temperature Strong Coupling of Au Nanorod-WSe$_2$ Heterostructures

**Jinxiu Wen**[1, 3], **Hao Wang**[1, 3], **Huanjun Chen**[1, 2], **Shaozhi Deng**[1, 2] **and Ningsheng Xu**[1, 2]

[1] State Key Laboratory of Optoelectronic Materials and Technologies, Guangdong Province Key Laboratory of Display Material and Technology, Sun Yat-sen University, Guangzhou 510275, China

[2] School of Electronics and Information Technology, Sun Yat-sen University, Guangzhou 510006, China

[3] School of Physics, Sun Yat-sen University, Guangzhou 510275, China

E-mail: chenhj8@mail.sysu.edu.cn and stsdsz@mail.sysu.edu.cn

**Abstract:**
All-solid-state strong light-matter coupling systems with large vacuum Rabi splitting are great important for quantum information application, such as quantum manipulation, quantum information storage and processing. The monolayer transition metal dichalcogenides (TMDs) have been explored as excellent candidates for the strong light-matter interaction, due to their extraordinary exciton binding energies and remarkable optical properties. Here, for both of experimental and theoretical aspects, we explored resonance coupling effect between exciton and plasmonic nanocavity in heterostructures consisting of monolayer tungsten diselenide (WSe$_2$) and an individual Au nanorod. We also study the influences on the resonance coupling of various parameters, including localized surface plasmon resonances of Au nanorods with varied topological aspects, separation between Au nanorod and monolayer WSe$_2$ surface, and the thickness of WSe$_2$. More importantly, the resonance coupling can approach the strong coupling regime at room-temperature by selecting appropriate parameters, where an anti-crossing behavior with the vacuum Rabi splitting strength of 98 meV was observed on the energy diagram.

## 1. Introduction

Understanding and controlling the strong light-matter coupling are great promising for quantum optical application [1–3]. The interaction of the quantum emitters can be modified by its surrounding local electromagnetic environment. This means that the spontaneous emission rate of emitter can be modified by optical cavity. If the emitter and optical cavity interaction strength become large than their individual dissipation, the strong coupling regime is achieved [4, 5]. In this case, strong coupling results in the formation of mixed energy states with part-light and part-matter and anti-crossing behavior can be observed in the optical spectra, characterized by the vacuum Rabi splitting (VRS). In the field of quantum information science, achieving strong coupling effect or, in other words, quantum coherent oscillation between the emitter and optical cavity is the prerequisite for quantum manipulation, quantum information storage and processing [6–8]. The previous reports of strong coupling systems have been demonstrated, such as atoms-microcavity [9], quantum dots-photonic crystals [5, 10], quantum wells-distributed Bragg reflectors (DBR) cavities [11] as well as the later development of organic molecular-plasmonic nanocavity [12–14]. Whereas these systems were performed either non-solid-sate or suffering many experiment obstacles, like ultrahigh vacuum, cryogenic temperature, complex fabrication procedures and highly unstable. At this stage, the exploration of novel all-solid-state systems suitable for CMOS-compatible optoelectronic devices at room-temperature is strongly desired.

The layered transition metal dichalcogenides (TMDs), so far over 40 different TMDs found with different compounds, have attracted increasing attention in recent years. The remarkable optical and electronic properties of monolayer TMDs have been discovered, including the intriguing electronic structure owing to the breaking of inversion symmetry and strong spin-valley coupling, the indirect-to-direct band-gap semiconductors with strong light emission ranging from the near-infrared to the visible, and particularly very large exciton binding energies (0.3 eV ~ 0.9 eV) due to the geometrical confinement and weak dielectric screening. With these properties, it has motivated extensive studies to explore its potential physical phenomena, especially on strong light-matter interaction [15–19]. During the past two years, a few studies about all-solid-state strong coupling systems have been reported for TMDs and optical cavities, including TMDs interaction with DBR cavities [20–22] or plasmonic of metal nanostructure arrays [23, 24]. However, to the best our knowledge, the monolayer TMDs-DBR cavity systems occasionally need to be performed in cryogenic temperature to observe the distinct Rabi splitting phenomenon. On the other hand, the designed structures like DBR and metal nanostructure arrays are usually fabricated complex and difficult reproducibility of experiments, which limits the development of active-controlled optoelectronic devices in the future.

Here, we for the first time proposal the observation of strong coupling in all-solid-state system composed of monolayer $WSe_2$ and an individual Au nanorod at room temperature. The monolayer $WSe_2$ exciton transition efficiency is stronger at room temperature than the cryogenic temperature in near-infrared regime, suggesting the exciton-optical interaction suitability for room temperature [25]. Furthermore, the Au nanorod can confined the electromagnetic fields into an ultrasmall volume that strongly enhance the interaction with excitons. Whereby both of these merits can greatly in favor of the two-dimensional $WSe_2$ exciton and Au nanorod plasmon entering the strong coupling regime, where hybrid modes are formed and give rise to an anti-crossing behavior with the VRS of 98 meV on the energy diagram.

## 2. Results and discussion

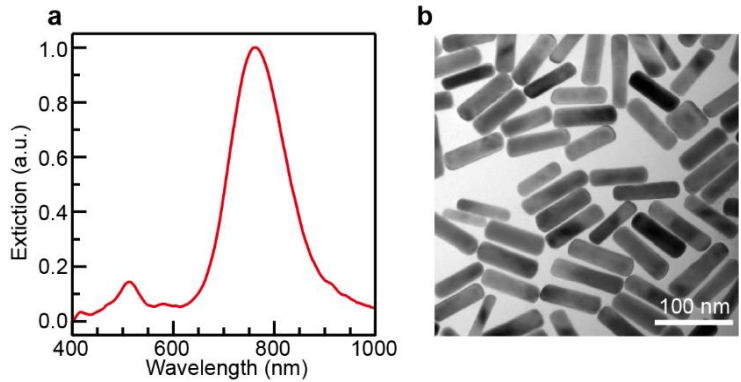

**Figure 1.** Characterization of Au nanorods sample. (a) Extinction spectrum of the pristine Au nanorod sample. (b) TEM image of the pristine Au nanorods.

The most intriguing characteristic of Au nanorods as plasmonic nanocavity is their localized surface plasmon resonances (LSPRs). The electromagnetic fields associated with the LSPRs mode are tightly confined to deep sub-wavelength volumes in all three dimensions which provides a way to focus light in the nanoscale and large electric enhancements around the Au nanorods. Furthermore, the Au nanorods with very high yields and uniformity can be easily synthesized using the mature methods and the LSPRs wavelength can be facile tuned by tailoring their topological aspects at ambient conditions [26]. Usually elongated Au nanorods exhibit two types of plasmon resonances, the transverse plasmon mode (TPM) associated with electron oscillations along the diameter direction and the longitudinal plasmon mode (LPM) owing to electron oscillations along the length direction. In our study, we focused on the LPM, which was strong dependent on the aspect ratios of the Au nanorods. The Au nanorods were synthesized using the seed-mediated method [27] in our study. Figure 1(a) gives the extinction spectrum of pristine Au nanorod sample by adding 20 μL seeds which exhibited LPM wavelength of 766 nm. The morphologies of Au nanorod can be seen clearly from the TEM image which showed uniform size and shape distribute (figure 1(b)). The average length, diameter, and aspect ratio of the pristine Au nanorods are 80 ± 6 nm, 23 ± 2 nm, 3.4 ± 0.4. The monolayer $WSe_2$ flake was grown by chemical vapor deposition (CVD) method. Figure 2(a) exhibits triangular shape with edge lengths ~50 μm. The monolayer nature of $WSe_2$ flake can be confirmed by typical atomic force microscopy (AFM) image and the thickness was determined to be ~ 0.9 nm (figure 2(b)). Figure 2(c) shows the high-resolution transmission electron microscopy (HRTEM) image of the single-crystalline $WSe_2$ flake which clearly resolves the atomic lattice and the selective area electron diffraction (SAED) pattern (the inset of figure 2(c)) shows a hexagonal crystal structure. The absence of the $B_{2g}^1$ peak at ~308 cm$^{-1}$ indicated that the flake was monolayer (Figure 2(d)), which is in good agreement with previous Raman studies [28]. Both of the structure and light emission of monolayer $WSe_2$ flake are very uniform, which can be manifested from the Raman intensity and photoluminescence position mapping in figure 2(e) and (f).

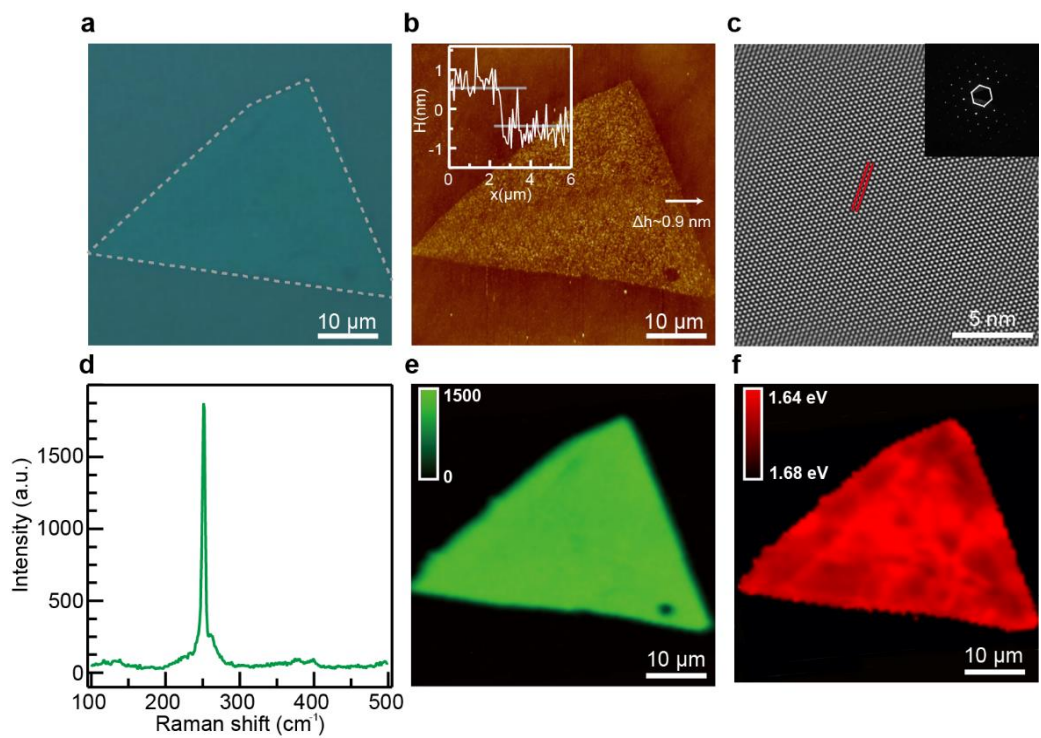

**Figure 2.** Characterization of monolayer $WSe_2$ flake. (a) Optical image of the monolayer $WSe_2$. (b)

AFM image of the monolayer WSe$_2$ flake showing the thickness of ~ 0.9 nm. (c) High-resolution transmission electron microscopy (HRTEM) image of the monolayer WSe$_2$. Insert showing the selective area electron diffraction (SAED) pattern, which shows a hexagonal crystal structure. (d) Raman spectrum of triangular WSe$_2$. (e), (f) photoluminescence position and Raman intensity mapping of WSe$_2$ flake corresponding to (e), respectively.

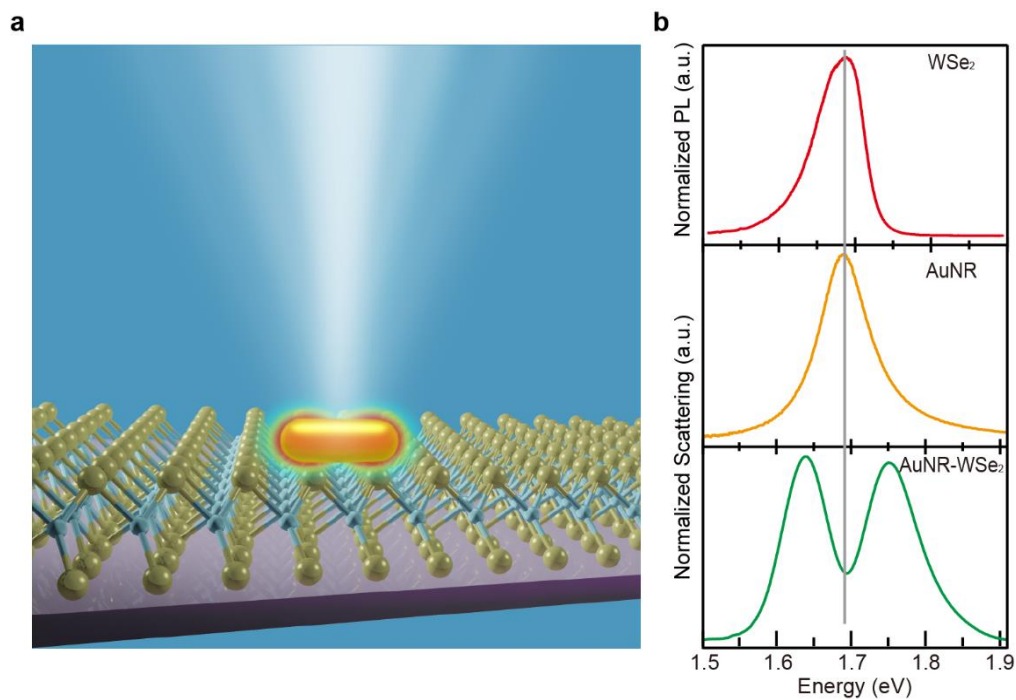

**Figure 3.** Au nanorod−WSe$_2$ heterostructures. (a) Schematic showing the hybrid structure composed of an individual Au nanorods coupled to a monolayer WSe$_2$ flake. (b) Photoluminescence spectrum of the monolayer WSe$_2$ (upper), scattering spectrum of an individual Au nanorod (middle), and scattering spectrum of Au nanorod−WSe$_2$ heterostructure (lower).

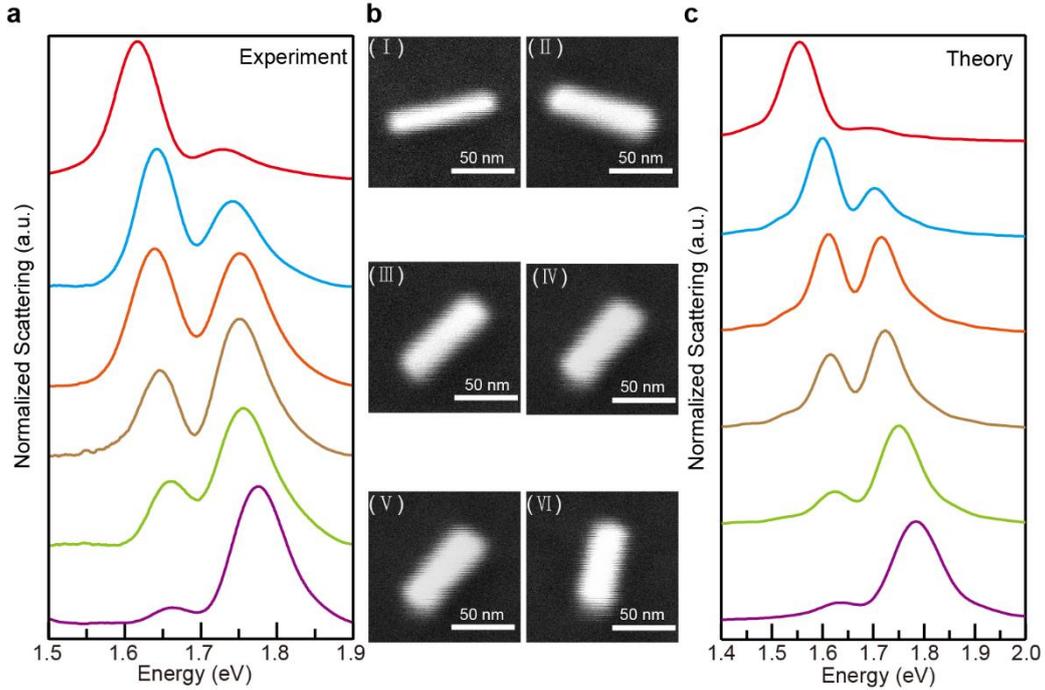

**Figure 4.** Strong coupling in Au nanorod−WSe$_2$ heterostructures. (a) Dark-field scattering spectra from different individual Au nanorods coupled to the same monolayer WSe$_2$. (b) SEM images of Au nanorods coupled to monolayer WSe$_2$, which correspond to the scattering spectra shown in (a) (from top to bottom). The aspect ratios of the Au nanorods from (I) to (VI) are 4.4, 3.6, 3.5, 3.4, 3.2 and 3.0, respectively. (c) Calculated spectra corresponding to the experimental spectra shown in (a).

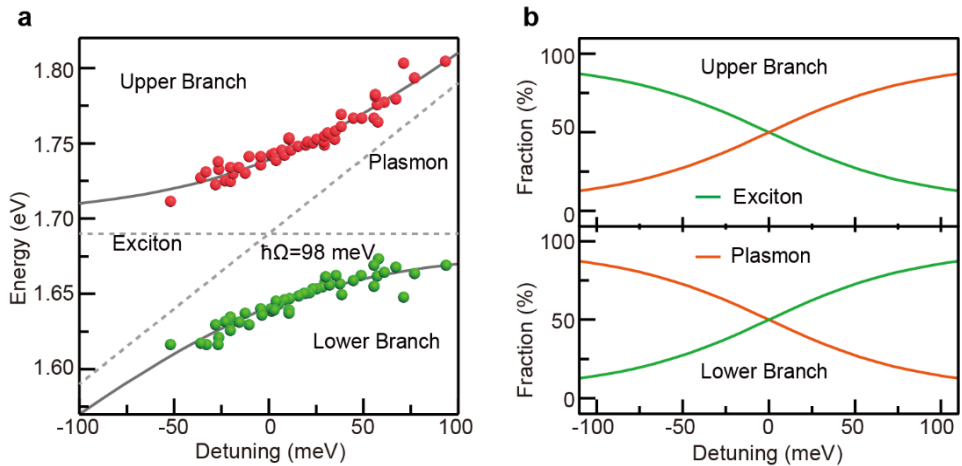

**Figure 5.** Anti-crossing behavior of Au nanorod−WSe$_2$ heterostructures. (a) Normalized scattering energy diagrams of varied detunings between the exciton and plasmon resonance coupling of the heterostructures, showing an anti-crossing behavior between the upper and lower branches. The two energy branches are fitted to a coupled oscillator model, showing Rabi splitting of 98 meV. (b) Exciton and plasmon fractions of the upper branch and lower branch of Au nanorod−WSe$_2$ heterostructures, respectively.

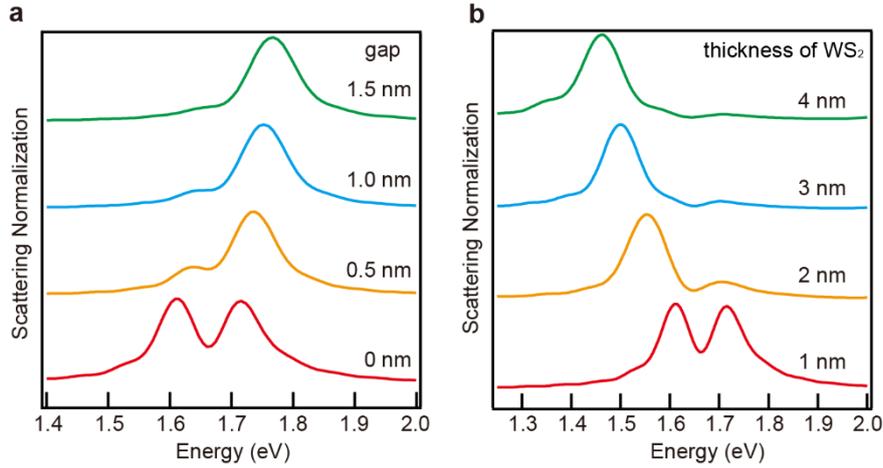

**Figure 6.** Calculaed scattering spectra of Au nanorod−WSe$_2$ heterostructures using FDTD method. The aspect ratio of the Au nanorod is kept as 3.5 (a) Dependence of the strong coupling on the separations between the Au nanorod and WSe$_2$ surface, while the thickness of the WSe$_2$ is fixed at 1 nm. (b) Dependence of the strong coupling on the thickness of WSe$_2$ with the Au nanorod contact to the WSe$_2$ surface.

The strong light-matter coupling hybrid nanostructure consists of the Au nanorod depositing onto the monolayer WSe$_2$ flake (figure 3(a)). Various individual Au nanorods were sparsely distributed onto the WSe$_2$ flake and SiO$_2$ substrate, which were allowed for identifying their scattering spectra using dark-field scattering spectroscopy and morphologies using the scanning electron microscope (SEM) technique. The pristine monolayer WSe$_2$ flake exhibits a very sharp and strong exciton emission centering at 1.67 eV, with a narrow line width ($\gamma_{ex}$) of 47 meV (figure 3(b) upper panel). The dark-field scattering spectra of individual Au nanorod depositing onto SiO$_2$ and monolayer WSe$_2$ flake were measured respectively (Figure 3(b) middle and lower panel). The typical Au nanorod scattering spectrum exhibits LPM peak at 1.67 eV and the plasmon line width ($\gamma_{pl}$) to be estimated as 82 meV, which is resonance with the exciton emission of WSe$_2$ flake. Whereas the peak splits into two distinctly scattering peaks from the Au nanorod coupling with the monolayer WSe$_2$ flake, where the high-energy (HEM) and low-energy (LEM) hybrid modes are formed (Figure 3(b) lower panel). In order thorough demonstrate the resonance coupling of heterostructure, the dark-field scattering spectra were measured from various individual Au nanorods depositing onto the same WSe$_2$ flake that of different detuning energies between the plasmon frequencies and exciton transition. To this end, seven representative scattering spectra of Au nanorod−WSe$_2$ heterostructure shown in figure 4(a) were performed the corresponding SEM characterization to determine their morphologies (figure 4(b)). The LPM frequencies were decreased from bottom to top with the aspect ratio increasing from 3.0 to 4.4. The intensities of the scattering spectra obtained have been normalized. When LPM frequencies are on the high-energy side away from the exciton transition, the HEM dominates and redshift with decreasing LPM frequencies. As the LPM frequencies become overlapped with the exciton transition, the HEM and LEM are comparable with each other. When the LPM frequencies move further from the exciton to the low-energy side, the LEM higher than the HEM and redshifts as the detuning energy becomes larger. To further understanding the strong coupling effect of the Au nanorod−WSe$_2$ heterostructure, we carried out the finite-difference time-domain (FDTD) method to calculate the light scattering properties of heterostructures. The calculated scattering spectra (figure 4(c)) shapes and evolvements manifest similar with the experimental results. To further confirm that the heterostructure indeed approach the strong coupling regime, the anti-crossing behavior is the one of the essential feature of the strong coupling effect between exciton transition and optical mode. Whereas such phenomenon can be observed on the normalized scattering spectra evolution of the Au nanorod−WSe$_2$ heterostructures by continue tuning the energy detuning of LPM energy across the exciton transition energy. Two distinct branches associated with LEM and HEM respectively can be seen in the coupling energy diagram, namely the upper branch and lower branch, which manifests a clearly anti-crossing behavior in figure 5(a). Accordingly, a coupled harmonic oscillator model was then adopted to describe the strong coupling effect, which are excellent agreed with the experimental data [13]. As shown in figure 5(a), the Rabi splitting energy, $\hbar\Omega$, can be seen to be 98 meV at zero energy detuning. On the other hand, the second condition of strong coupling is that a Rabi splitting energy should be fulfill:

($\hbar\Omega > \frac{\gamma_{pl} + \gamma_{ex}}{2}$) [20], which is also satisfied in our results. Therefore, the resonance interaction between individual Au nanorod and monolayer WSe$_2$ enters the strong coupling regime. In this regime, two new hybridized stated with part-light and part-matter will be formed due to the coherent energy exchange between the exciton and plasmon resonance. We also utilized the coupled harmonic oscillator model to calculate the respective contributions from exciton and plasmon components for upper and lower branches. As shown in figure 5(b), by detuning the plasmon resonance to low-energy side of the exciton transition, the lower branch is more plasmon-like while the high-energy one exciton-like, and vice versa for detuning to the high-energy side. Furthermore, the dependence of strong coupling on separation between Au nanorod and monolayer WSe$_2$ surface, as well as the thickness of WSe$_2$ were considered in our calculation (figure 6(a) and (b)). As the Au nanorod is located far away from the WSe$_2$ surface, the split LEM will vanish quickly and disappear for separation larger than 1.5 nm (figure 6(a)). On the other hand, the exciton emission efficiency is the strongest for monolayer WSe$_2$ and decrease as the number of WSe$_2$ flakes increase, so that the intensities of HEM tend to diminish when the thickness of WSe$_2$ become increasing (figure 6(b)). These results were manifested the strong coupling with two distinct split modes when the Au nanorod and monolayer WSe$_2$ contacted closely, which are excellent consistent with our experiment.

## 3. Conclusion

In summary, we for the first time observed the strong coupling in exciton-plasmonic nanocavities heterostructures composed of an individual Au nanorod and monolayer WSe$_2$ at room-temperature. The resonance interaction was investigated by the single-particle dark-field scattering spectroscopy, FDTD simulations, and the coupled harmonic oscillator model. An excellent agreement between the experimental and theoretical results have been obtained. The coupling strength of Au nanorod−WSe$_2$ heterostructures can be facile tuned by selecting various topological aspects of Au nanorods, where an anti-crossing behavior is observed on the energy diagram with the giant vacuum Rabi splitting of 98 meV. Moreover, we believe that the proposed Au nanorod−WSe$_2$ heterostructures can pave the way to explore the strong coupling of all-solid-state systems in near-infrared regime that will widely utilized in future quantum optical and quantum information application.

## 4. Methods

### 4.1. Characterizations

The scattering spectra of individual Au nanorods over the monolayer WSe$_2$ were recorded on a dark-field optical microscope (Olympus BX51) that was integrated with a quartz-tungsten-halogen lamp (100 W), a monochromator (Acton SpectraPro 2360), and a charge-coupled device camera (Princeton Instruments Pixis 400BR_eXcelon). The camera was thermoelectrically cooled to -70 °C during the measurements. A dark-field objective (100×, numerical aperture 0.80) was employed for both illuminating the nanocrystals with the white excitation light and collecting the scattered light.

For measuring the extinction spectra of Au nanorods colloidal samples, a HITACHI U-4100 UV/visible/near-infrared spectrophotometer with an incidence spot size of 5 mm was utilized. The Raman and PL spectra of WSe$_2$ monolayer were collected using a Renishaw inVia Reflex system with a dark-field microscopy (Leica). The excitation laser of 532 nm was focused onto the samples with a diameter of ~1 μm through a 50× objective (NA=0.8). SEM images of individual Au nanorods were acquired using an FEI Quanta 450 microscope. The thickness of monolayer WSe$_2$ was measured using AFM (NTEGRA Spectra). HRTEM and SAED measurements were conducted on the same field emission TEM (FEI Tecnai[3] G2 60-300) with operating at 300 kV.

### 4.2. FDTD Calculations

Finite-difference time-domain (FDTD) method was utilized to calculate the optical properties of the Au nanorod−WSe$_2$ heterostructure. The Au nanorod was modeled as a cylinder capped with a hemisphere at each end, and was placed on a WSe$_2$ layer on top of a 300 nm thick SiO$_2$ layer. The bulk dielectric function of gold was used [29]. The dielectric function of the monolayer WSe$_2$ was adopted from previous reported values [30]. In this manner, the exciton energy is 1.66 eV. A dielectric constant of 2.25 was used for the SiO$_2$ substrate. For the calculation of scattering spectra of Au nanorod–WSe$_2$ heterostructure, the lengths and the diameters of the Au nanorods set to (78 nm, 26 nm), (80 nm, 25 nm), (81 nm, 24 nm), (83 nm, 24 nm), (86nm, 24 nm) and (87 nm, 20 nm) corresponding the aspect ratio from 3.0 to 4.4. Optimization calculations were carried out to make the Au nanorod contact with the WSe$_2$ and thickness (1 nm) of the WSe$_2$ monolayer.

### 4.3. Coupled Harmonic Oscillator Model

The hybridized states can be described using a coupled harmonic oscillator model as:

$$\begin{pmatrix} E_{pl} + i\dfrac{\gamma_{pl}}{2} & g \\ g & E_{ex} + i\dfrac{\gamma_{ex}}{2} \end{pmatrix} \begin{pmatrix} \alpha \\ \beta \end{pmatrix} = E_{\pm} \begin{pmatrix} \alpha \\ \beta \end{pmatrix} \quad (1)$$

where $E_{pl}$, $E_{ex}$ and $\gamma_{pl}$, $\gamma_{ex}$ are the energies of the uncoupled plasmon and exciton transition and the dissipation rates respectively. g is the coupling rate, $E_{\pm}$ are the eigenvalues of the coupled system which describe the energies of the hybridized states, and α and β are the coefficients of the liner combination of the plasmon and exciton, where $|\alpha^2+\beta^2| = 1$. To simplify the analysis we ignore the dissipations, and the eigenvalues are:

$$E_{\pm} = \dfrac{E_{pl} + E_{ex}}{2} \pm \dfrac{\sqrt{4g^2 + \delta^2}}{2} \quad (2)$$

where $\delta = E_{pl} - E_{ex}$ is the detuning between the plasmon resonance and exciton transition energy. The Rabi splitting energy, $\hbar\Omega = 2g$, can be obtained when $E_{pl} = E_{ex}$. Besides, according to equation (2),

$$g = \sqrt{(E_+ - E_{ex})(E_{ex} - E_-)} \tag{3}$$

## Acknowledgments

This work was financially supported by the National Natural Science Foundation of China (Grant Nos. 51290271, 11474364), the National Key Basic Research Program of China (Grant Nos. 2013CB933601, 2013YQ12034506), the Guangdong Natural Science Funds for Distinguished Young Scholars (Grant No. 2014A030306017), Pearl River S&T Nova Program of Guangzhou (Grant Nos. 201610010084), and the Guangdong Special Support Program.

## References

[1] Matsukevich D N and Kuzmich A 2004 Quantum state transfer between matter and light. *Science*, **306** 663–666
[2] Chen W, Beck K M, Bücker R, Gullans M, Lukin M D, Tanji-Suzuki H, and Vuletić V 2013 All-optical switch and transistor gated by one stored photon *Science* **341** 768–770
[3] Sanvitto D, and Kéna-Cohen S 2015 The road towards polaritonic devices *Nat. Mater.* **15** 1061–1073
[4] Koenderink A F, Alù A and Polman A 2015 Nanophotonics: Shrinking light-based technology *Science* **348** 516–521
[5] Yoshie T, Scherer A, Hendrickson J, Khitrova G, Gibbs H M, Rupper G, Ell C, Shchekin O B and Deppe D G 2004 Vacuum Rabi splitting with a single quantum dot in a photonic crystal nanocavity *Nature* **432** 200–203
[6] Mabuchi H and Doherty A C 2002 Cavity quantum electrodynamics: coherence in context *Science* **298** 1372–1377
[7] Sillanpää M A, Park J I and Simmonds R W 2007 Coherent quantum state storage and transfer between two phase qubits via a resonant cavity *Nature* **449** 438–442
[8] Kimble H J 2008 The Quantum Internet *Nature* **453** 1023–1030
[9] Hood C J, Lynn T W, Doherty A C, Parkins A S, and Kimble H J 2000 The atom-cavity microscope: single atoms bound in orbit by single photons *Science* **287** 1447–1453
[10] Reithmaier J R *et al* 2004 Strong coupling in a single quantum dot semiconductor microcavity system *Nature* **432** 197–200
[11] Khitrova G, Gibbs H M, Kira M, Koch S W and Scherer A 2006 Vacuum Rabi splitting in semiconductors *Nat. Phys.* **2** 81–90
[12] Fofang N T, Park T H, Neumann O, Mirin N A, Nordlander P and Halas N J 2008 Plexcitonic nanoparticles: plasmon exciton coupling in nanoshell– J-aggregate complexes *Nano Lett.* **8** 3481–3487
[13] Zengin G, Wersäll M, Nilsson S, Antosiewicz T J, Käll M and Shegai T 2015 Realizing strong light-matter interactions between single-nanoparticle plasmons and molecular excitons at ambient conditions *Phys. Rev. Lett.* **114** 157401
[14] Santhosh K, Bitton O, Chuntonov L and Haran G 2016 Vacuum Rabi splitting in a plasmonic cavity at the single quantum emitter limit *Nat. Commun.* **7** 11823

[15] Zhang H 2015 Ultrathin two-dimensional nanomaterials. *ACS Nano* **9** 9451–9469

[16] Zhou K G and Zhang H L 2015 Lighten the olympia of the flatland: probing and manipulating the photonic properties of 2D transition-metal dichalcogenides *Small* **11** 3206–3220

[17] Mak K F and Shan J 2016 Photonics and optoelectronics of 2D semiconductor transition metal dichalcogenides *Nat. Photonics* **10** 216–226

[18] Ramasubramaniam A 2012 Large excitonic effects in monolayers of molybdenum and tungsten dichalcogenides *Phys. Rev. B* **86** 115409

[19] He K, Kumar N, Zhao L, Wang Z, Mak K F, Zhao H and Shan J 2014 Tightly bound excitons in monolayer $WSe_2$. *Phys. Rev. Lett.* **113** 026803

[20] Dufferwiel S *et al* 2015 Exciton-polaritons in van der Waals heterostructures embedded in tunable microcavities *Nat. Commun.* **6** 8579

[21] Liu X, Galfsky T, Sun Z, Xia F, Lin E C, Lee Y H and Menon V M 2015 Strong light matter coupling in two-dimensional atomic crystals *Nat. Photonics* **9** 30–34

[22] Lundt N *et al* 2016 Room-temperature Tamm-plasmon exciton-polaritons with a $WSe_2$ monolayer *Nat. Commun.* **7** 13328

[23] Liu W, Lee B, Naylor C H, Ee H S, Park J, Johnson A C and Agarwal R 2016 Strong exciton plasmon coupling in $MoS_2$ coupled with plasmonic lattice *Nano Lett.* **16** 1262–1269

[24] Wang S J *et al* 2016 Coherent coupling of $WS_2$ monolayers with metallic photonic nanostructures at room temperature *Nano Lett.* **16,** 4368–4374

[25] Zhang X X, You Y, Zhao S Y F and Heinz T F 2015 Experimental evidence for dark excitons in monolayer $WSe_2$ *Phys. Rev. Lett.* **115** 257403

[26] Ni W H, Kou X S, Yang Z and Wang J F 2008 Tailoring longitudinal surface plasmon wavelengths, scattering and absorption cross sections of gold nanorods *ACS Nano* **2** 677–686

[27] Chen H J, Shao L, Li Q and Wang J F 2013 Gold nanorods and their plasmonic properties *Chem. Soc. Rev.* **42** 2679–2724

[28] Zhao W, Ghorannevis Z, Amara K K, Pang J R, Toh M, Zhang X, Kloc C, Tan P H and Eda G 2013 Lattice dynamics in mono-and few-layer sheets of $WS_2$ and $WSe_2$ *Nanoscale* **5** 9677–9683

[29] Johnson P B and Christy R W 1972 Optical constants of the noble metals *Phys. Rev. B* **6** 4370

[30] Li, Y. et al. 2014 Measurement of the optical dielectric function of monolayer transition-metal dichalcogenides: $MoS_2$, $MoSe_2$, $WS_2$, and $WSe_2$ *Phys. Rev. B* **90** 205422